\documentclass{aa}
\usepackage{natbib}
\usepackage{graphicx}
\usepackage{amssymb}
\usepackage{multirow}
\usepackage[varg]{txfonts}

\bibpunct{(}{)}{;}{a}{}{,}

\begin{document}

\title{Mild evolution of the stellar metallicity gradients of disc galaxies}

\author{Patricia B. Tissera\inst{1,2}
 \and Rubens E. G. Machado\inst{1,3,4} 
 \and Jos\'e M. Vilchez\inst{5}
 \and Susana E. Pedrosa\inst{6}
\and Patricia Sanchez-Blazquez\inst{7,8}
\and Silvio Varela\inst{1}}
\offprints{P. B. Tissera}
\institute{Departamento de Ciencias Fisicas, Universidad Andres Bello, Fernandez Concha 700, Santiago, Chile.\\
\email{patricia.tissera@unab.cl}
\and Millennium Institute of Astrophysics, Av. Republica 220, Santiago, Chile.
\and Universidade Federal de Ouro Preto, Departamento de F\'isica, Campus Universitario Morro do Cruzeiro, 35400-000, Ouro Preto, Brazil.
\and Departamento Acad\^emico de F\'isica, Universidade Tecnol\'ogica Federal do
Paran\'a, Rua Sete de Setembro 3165, 80230-901 Curitiba, Brazil.
\and Instituto de Astrof\'isica de Andaluc\'ia (CSIC), Glorieta de la Astronom\'ia s/n, E-18008 Granada, Spain.
\and Instituto de Astronom\'{\i}a y F\'{\i}sica del Espacio,
CONICET-UBA, Casilla de Correos 67, Suc. 28, C1428ZAA, Ciudad
Aut\'onoma de Buenos Aires, Argentina.
\and Departamento de F\'isica Te\'orica, Universidad Aut\'onoma de Madrid, 28049 Madrid, Spain.
\and  Instituto de Astrof\'isica , Universidad Pontifica Cat\'olica de Chile, Av. Vicu\~na Mackenna 4860, Santiago, Chile.
}

\date{Received / Accepted}

\abstract
{The metallicity gradients of the stellar populations in disc galaxies and their evolution store relevant information on the disc formation history and on those processes which could mix stars {\it a posteriori}, such as migration, bars and/or galaxy-galaxy interactions. }
{{We aim to investigate the evolution of the metallicity gradients of the whole stellar populations in disc components of simulated galaxies in a cosmological context.}}
{We analyse simulated disc galaxies selected from a cosmological hydrodynamical simulation that includes chemical evolution and a physically motivated Supernova feedback capable of driving mass-loaded galactic winds.
}
{We detect a mild evolution  with redshift in the metallicity slopes of $-0.02 \pm 0.01$ dex~kpc$^{-1}$ from $z\sim 1$.  If the metallicity profiles are normalised by the effective radius of the stellar disc, the slopes show no clear evolution for $z < 1$, with a median value of approximately $-0.23$ dex ~$r_{\rm reff}^{-1}$.  As a function of stellar mass, we find  that metallicity gradients steepen for stellar masses smaller  than $\sim 10^{10.3} {\rm M_{\odot}}$ while the trend reverses for higher stellar masses, in the redshift range $z=[0,1]$. Galaxies with small stellar masses  have discs with larger $r_{\rm reff}$ and flatter metallicity gradients than expected. We detect  migration albeit weaker than in previous works. }
{Our stellar discs show a mild evolution of the stellar metallicity slopes up to $z\sim 1,$ which is well-matched by the evolution calculated  archeologically from the abundance distributions of mono-age stellar populations at $z\sim 0$.  The dispersion in the relations allows for stronger individual evolutions.
  Overall, Supernova feedback could explain the trends but an impact of migration can not be totally discarded.
Galaxy-galaxy interactions or small satellite accretions can also contribute to modify the metallicity profiles in the outer parts. Disentangling the effects of these processes for individual galaxies  is still a challenge
in a cosmological context.
  } 


\titlerunning{Evolution of the stellar metallicity gradients}
\authorrunning{Tissera et al.}
\maketitle

%%%%%%%%%%%%%%%%%%%%%%%%%%%%%%%%%%%%%%%%%%%%%%%%%%%%%%%%%%%
\section{Introduction}

The chemical abundances of the stellar populations (SPs) in galaxies of different morphologies provide important clues to understand their history of formation. Chemical patterns store relevant information on the properties of the interstellar medium (ISM)  and on the physical processes that might modify them along the galaxy evolutionary paths \citep[e.g.][]{matteucci1986,molla1997,chiappini2001,pipi2008,gibson2013, tissera2016a}. In an inside-out  scenario for disc formation, gas-phase metallicity profiles with negative gradients are a natural outcome if star formation is driven mainly by the gas  density in the discs. Supernova (SN) feedback can regulate the star formation activity and modulate the metallicity content of galaxies if mass-loaded galactic winds are triggered \citep{gibson2013,tissera2016a}. New-born stars acquire the chemical abundances of  the ISM from where they formed. Later on, the stellar abundance distributions could be affected by dynamical processes that redistribute stars in the disc components such as bars, migration \citep[e.g.][]{minchev2014}, or galaxy-galaxy interactions and mergers \citep[e.g.][]{dimatteo2009, perez2011}.

In the Local Universe, most of our current knowledge of chemical distributions in disc galaxies comes from  studies of  the ISM via observations of HII regions or  young stellar populations. Current observations are consistent with the metallicity profiles of the ISM having negative metallicity gradients, on average. Metallicity gradients in units of dex${~\rm kpc^{-1}}$ determine tight correlations with the global properties of galaxies such as the stellar mass or the size,  which are erased when normalised by a characteristics radius such as the half-mass radius \citep[e.g.][]{zaritsky1994,sanchez2013Califa}. There are few indirect  estimations of the evolution of the stellar metallicity gradients using planetary nebulae (PN) \citep[e.g.][]{henry2010,stang2010}. \citet{maciel2003} calculated the gradients of SPs with different ages in the Milky Way disc, finding a signal of increasingly negative metallicity gradients for older stars. Recently \citet{magrini2016} analysed the metallicity gradients determined by PN for four nearby galaxies, finding them to be flatter than those detected using HII regions. 
 Surveys such as {\small CALIFA} provide detailed information on the properties of the ISM and the  SPs, including their chemical abundances and age distributions on a variety of galaxies  \citep{sanchezb2014, gonzalezdelgado2015}.  The SDDS-IV MaNGA survey \citep{bundy2015} also investigate spatially resolved SPs and radial age and metallicity gradients for nearby galaxies \citep{li2015, wilkinson2015} and will provide a large statistical sample to confront with models. The available observations of HII regions for high-z galaxies do not allow the formation of a robust conclusion on the evolution of the metallicity gradients \citep[e.g.][]{yuan2011,queyrel2012,stott2014, jones2015}. In fact, high-redshift observations show a complex situation with gas-phase components showing a variety of metallicity gradients that could respond to the action
of different physical processes \citep[e.g.][]{cresci2010}.

A large effort has been invested in understanding the chemical patterns of galaxies using analytical and numerical chemical modelling \citep[e.g][]{brooks2007,calura2012,molla2015}. In particular, hydrodynamical simulations provide the chemical enrichment of baryons as galaxies are assembled in a cosmological framework, opening the possibility of understanding the interplay of different physical processes  in the non-linear regime of evolution \citep[e.g.][]{mosconi2001,lia2002,wier2009}. \citet{pilkington2012} carried out a comparison of the metallicity gradients in the ISM (traced by young SPs) obtained by different numerical and analytical models.
\citet{gibson2013} analysed different models of SN feedback schemes reporting different evolution for the gas-phase metallicity gradients. No evolution was reported when an enhanced SN model was used. \citet{tissera2016a} studied the gas-phase metallicity gradients in discs  and the specific star formation of the galaxies.
 For gas-phase metallicities, the simulated galaxies showed a correlation with stellar mass,
which was erased when the metallicity gradients were renormalised by the effective radius, reproducing observations by \citet[][and references there in.]{ho2015}
 They also found indications of a correlation between the abundance slopes and the specific star formation rate (sSFR) in agreement with observational findings \citep{stott2014}.  
As a function of redshift, the metallicity gradients of the gas-phase disc components were found to be more negative at higher redshift, principally for lower stellar-mass galaxies. The fraction of galaxies with positive metallicity gradients increased with increasing redshift and were found to be associated to mergers and interactions.
The trends found for the gas-phase metallicity gradients by  \citet{tissera2016a} are relevant to the discussion of this paper because we are using the same set of simulated galaxies.

The evolution of the metallicity gradients of the SPs provides important and complementary information on the chemodynamical evolution of galaxies. This is not only determined by the formation of discs but also  by those physical mechanisms that can disturb gradients {\it  a posteriori,} such as bar formation, migration,
and galaxy  interactions. Galaxy interactions  are known observationally \citep{lambas2003,ellison2008,rupke2010,micheldansac2008} and numerically \citep{bh96,tissera2000,perez2006,perez2011} to be an efficient mechanism to trigger gas inflows 
that transport low-metallicity material to inner regions. As a consequence, the  gas-phase  metallicity profiles become flatter. These gas inflows can trigger significant starbursts, which produce a two-fold effect: the new-ejected enriched material  can enrich both  the central regions, and steepen the metallicity gradients, and the outer parts of the galaxies via galactic fountains \citep{perez2011,dimatteo2013}. The new-born stars might reflect these changes  in metallicity depending on the relative contribution of the different stellar populations \citep{perez2011}.
Migration can also contribute to flatten the stellar chemical profiles by mixing them and transporting  old stars to the outer regions. This process is likely to be related to the transient spiral arms \citep{sellwood2002} or the effects of galaxy
interactions, which might produce radial mixing \citep{quillen2009}. Recent results using high-resolution simulations reported a more clear detection of migration, which showed the displacement of
old stars towards the outer part of the discs when a bar structure formed \citep{grand2016}.  It is important to note that if galaxies form in a hierarchical clustering
scenario, there are several processes that can affect the mixing of chemical elements in the ISM from which stars are formed (e.g. gas inflows, gas outflows, mergers), and later on, the distribution of
the stellar populations (e.g. minor mergers, migration, bar formation). Disentangling their effects is still a complicate task. 

In this paper, we study the evolution of the metallicity gradients of the SPs in the disc components of the set of simulated galaxies  analysed by \citet{tissera2016a,tissera2016b} at $z\sim 0$. These authors reported  the normalised and non-normalised stellar metallicity gradients  to be in agreement with results obtained  for galaxies in the {\small CALIFA} survey by \citet{sanchezb2014}.
At $z\sim 0$, the simulated stellar discs were reported to have half-mass radii ($r_{\rm reff}$) in very good agreement with observations \citep{sanchezb2014,wel2014}. 
This is an important outcome of our models considering that the free parameters were not fine-tuned to achieve this trend.
Overall, the simulated disc galaxies formed mainly in an inside-out fashion, from gas which conserved its specific angular momentum content. And if the discs are not perturbed, negative metallicity profiles are naturally formed as expected.

 At $z\sim 0,$ \citet{tissera2016b} analysed the metallicity profiles of the stellar populations   of the same discs we use in this work and reported  a change in the slope
of the relation between metallicity gradients and stellar mass, resembling a U-shape. These authors point out that this feature is more significant in the old SPs (i.e. older than 6 Gyr) than the young stars (i.e. younger than 2 Gyr). Old stars are reported to  have slightly steeper metallicity gradients than young SPs, on average.  A similar trend can be seen in CALIFA's results from \citet{sanchezb2014} and  \citet[][]{gonzalezdelgado2015}. 
In this paper, we extend the analysis done by \citet{tissera2016b} to $z\sim 1$. We note that our  galaxy sample provides different histories of formation for each system in agreement with the current cosmological framework, for the same sub-grid physics.

This paper is organised as follows. Section 2 summarises the  characteristics of the simulation and the galaxy catalogue. Section 3 describes the analysis and  Conclusions  summarises the main results.

%%%%%%%%%%%%%%%%%%%%%%%%%%%%%%%%%%%%%%%%%%%%%%%%%%%%%%%%%%%
\section{Simulated galaxies}

We analysed the same set of disc galaxies studied by \citet{tissera2016b} and \citet{tissera2016a}. The details on the numerical code used and the determination of the properties of discs can be found in the mentioned works and in \citet{pedrosa2015}. Here we summarise the main aspects. The analysed galaxies were selected from a cosmological volume consistent with the $\Lambda$ Cold Dark Matter scenario (the so-called simulation S230D from the Fenix Project) with $\Omega_{\Lambda}=0.7$, $\Omega_{\rm m}=0.3$, $\Omega_{b}=0.04$, a normalisation of the power spectrum of $\sigma_{8}=0.9$ and $H_{0}= 100\,h \ {\rm km} \ {\rm s}^{-1}\ {\rm Mpc}^{-1}$, with $h=0.7$. The simulated volume represents a box of $14$ Mpc comoving side, resolved with $2 \times 230^3$ initial particles, achieving a mass resolution of $5.9\times 10^{6}\,{h^{-1}\,\rm M_{\odot}}$ and $9.1\times 10^{5}\,{h^{-1}\,\rm M_{\odot}}$ for the dark matter and initial gas particles, respectively. The gravitational softening is $0.7$~kpc. 

We apply a version of the code {\small GADGET-3} \citep{springel2005}, optimised for massive parallel simulations of highly inhomogeneous systems. This version includes treatments for metal-dependent radiative cooling, stochastic star formation (SF), chemical enrichment, and a multiphase model for the interstellar medium (ISM) and the supernova (SN) feedback scheme of \citet{scan06}. The SN feedback model was able to successfully trigger galactic mass-loaded winds without introducing mass-scale free parameters. 
The code considers energy feedback by Type II (SNII) and Type Ia (SNIa) Supernovae. 
The code also includes the chemical evolution model developed by \citet{mosconi2001}. SNII and SNIa contributed with chemical elements estimated by adopting the chemical yields of \citet{WW95} and \citet{iwamoto1999}, respectively.

We use the galaxy catalogue constructed by \citet{tissera2016a}. The spheroidal and disc components are separated by applying a dynamical criterium based on the angular momentum content and the binding energy
momentum component in the direction of the total angular momentum, and $J_{\rm z,max}(E)$ is the maximum $J_{\rm z}$ over all particles at a given binding energy $E$ 
\citep[see][for details on the procedure and conditions used]{tissera2012}. 
At all analysed redshifts, only discs defined with more than 2000 star particles are studied. A few systems were discarded when the morphology did not allow for a robust linear fit to  the metallicity profile, as in galaxies strongly perturbed by close interactions, for example. After the removal of such objects, the adopted resolution criterium yielded 37, 23, 20 and 17 galaxies, for redshifts $z \sim$ 0, 0.4, 0.95 and 1.3, respectively. We note that the number of discs in the samples decreased for increasing redshift since we are maintaining the requirement of 2000 star particles to select the systems to be analysed for all redshifts.

%%%%%%%%%%%%%%%%%%%%%%%%%%%%%%%%%%%%%%%%%%%%%%%%%%%%%%%%%%%
\section{Analysis}

\begin{figure}
\includegraphics[width=0.9\columnwidth]{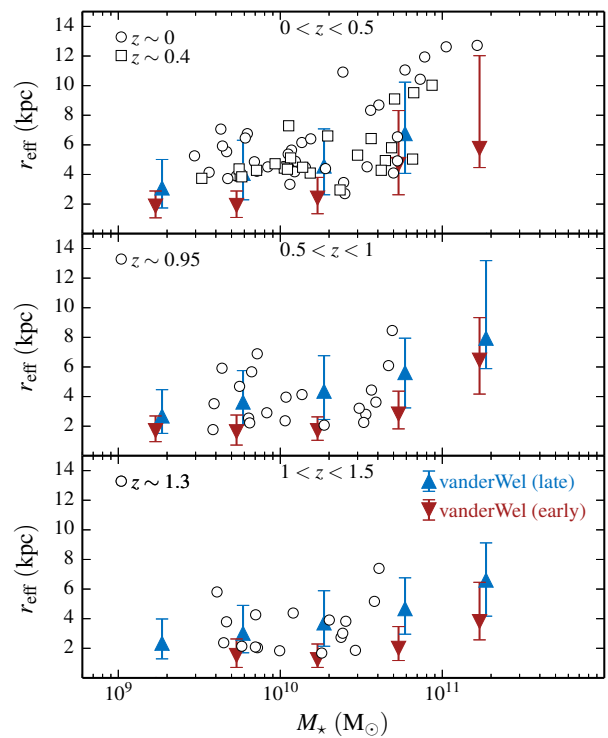}
\centering
\caption[]{Mean effective stellar radii for all simulated galaxies (open symbols) as a function of stellar mass for the four analysed redshifts. The observational data points of \citet{wel2014} in three redshift intervals are included for comparison and slightly offset horizontally for clarity.}
\label{fig:reffz}
\end{figure}

 Before analysing the evolution of  metallicity profiles, it is important to discuss if the sizes of  the simulated galaxies   evolve as a function of redshift as expected from observations.
At $z\sim 0$, \citet{tissera2016b} reported the $r_{\rm eff}$ of the stellar discs to be in good agreement with observations. Here, we  assess  to what extent our simulated $r_{\rm eff}$ are consistent with observations as a function of redshift. Figure~\ref{fig:reffz} shows the simulated stellar $r_{\rm eff}$ as a function of total stellar mass in the  four analysed redshift intervals and confronts them with observational estimations by \citet{wel2014}. As can be seen from this figure, the simulated galaxies have comparable $r_{\rm eff}$, which evolves similarly to the observed values estimated for late-type galaxies. We note that observations used the light-weighted distributions to estimate the  effective radius while we are considering the stellar mass distributions. For this comparison, the  $r_{\rm eff}$ are estimated using all stars in the main galaxies.  At a given stellar mass, there is a significant dispersion as can be appreciated from this figure, which is compatible with the observed values reported by \citet{wel2014}. Overall, we find that the decrease in size as a function of stellar mass with redshift is well-reproduced by our simulated discs in the analysed redshift range.

The disc sizes are determined by the angular momentum content of the infalling gas but they can also be affected by SN feedback and dynamical processes such as mergers   secular evolution and/or migration. \cite{pedrosa2015} found that galaxies in our sample  have stellar and gaseous discs formed consistently with the standard disc scenario based on  angular momentum conservation \citep{fall1980}. These simulated galaxies reproduce observed trends between the  specific angular momentum content, the potential well and the morphology  \citep{romanowsky2012}, and these relations are conserved as a function of redshift in this simulation. In the analysed redshift range, \citet{pedrosa2014} reported these galaxies to have not experienced major mergers since $z \sim 3$, which could have significantly disturbed the angular momentum content and the size of the systems.   Hence, for the analysed galaxies, the main path of disc formation is the infall of gas
with specific angular conservation. Then the gas is transformed into stars in the discs.
Afterwards, other mechanisms such  as migration, galaxy mergers, or interactions, can play a role in mixing stellar populations of different ages and chemical content \citep[e.g.][]{grand2016}.

%------------------------------------------------------------
%------------------------------------------------------------

%------------------------------------------------------------
\begin{figure*}
\includegraphics[width=2\columnwidth]{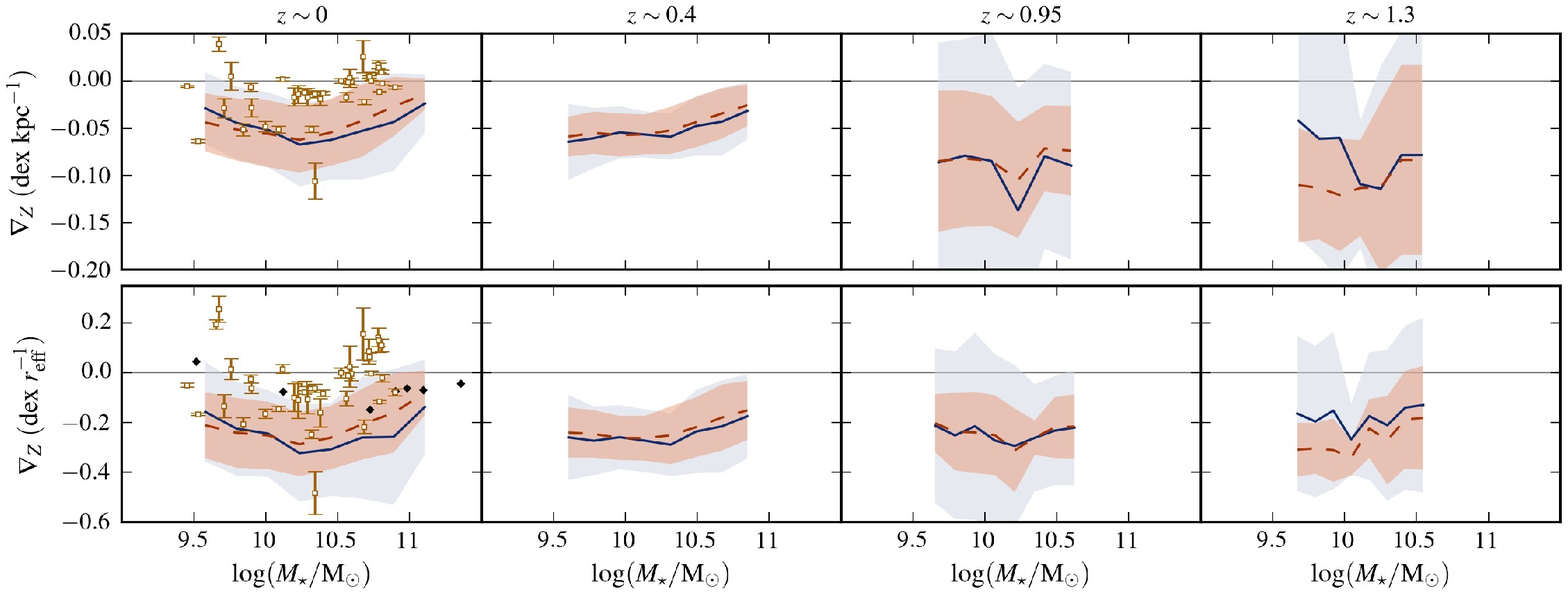}
\centering
\caption[]{Metallicity gradients as a function of stellar mass for the analysed disc galaxies at $z \sim 0$ (upper panels). 
The normalised gradients are displayed in the lower panels. Observational results for the metallicity gradients of the stellar discs in the CALIFA survey by \citet{sanchezb2014} (black diamonds) and \citet{gonzalezdelgado2015}  (yellow crosses, lower panel) are also displayed at $z \sim 0$ for comparison.  Mean estimations and their standard dispersion  in both 
radial intervals: $[0.5, 1.5]\,r_{\rm eff}$ (blue shades and lines) and $[0.5, 1.0]\,r_{\rm eff}$ (pink shades and lines) are included for  comparison. }
\label{fig:metzmasa}
\end{figure*}

\begin{figure*}
\includegraphics[width=2\columnwidth]{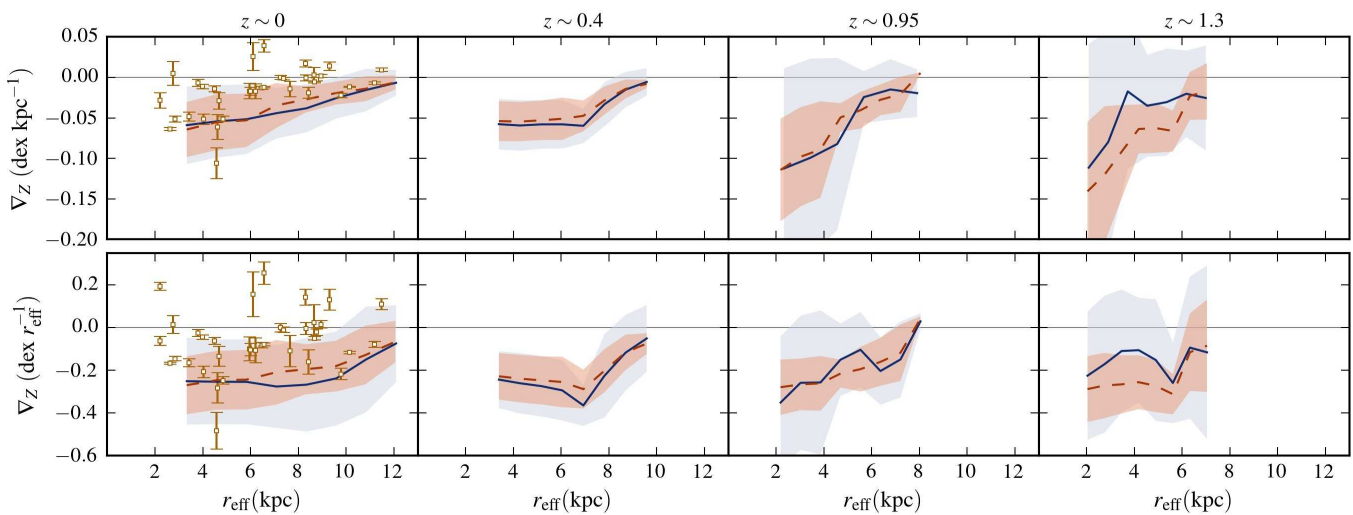}
\centering
\caption[]{Stellar metallicity gradients as a function of stellar $ r_{\rm eff} $  for the analysed  disc galaxies since $z \sim 1.3$ (upper panels). We also showed the 
distributions for the normalised stellar metallicity gradients at the same redshifts (lower panels). Standard deviations are included (shaded areas) as explained
in Fig.~\ref{fig:metzmasa}.}
\label{fig:metzradio}
\end{figure*}

\begin{figure}
\includegraphics[width=1.0\columnwidth]{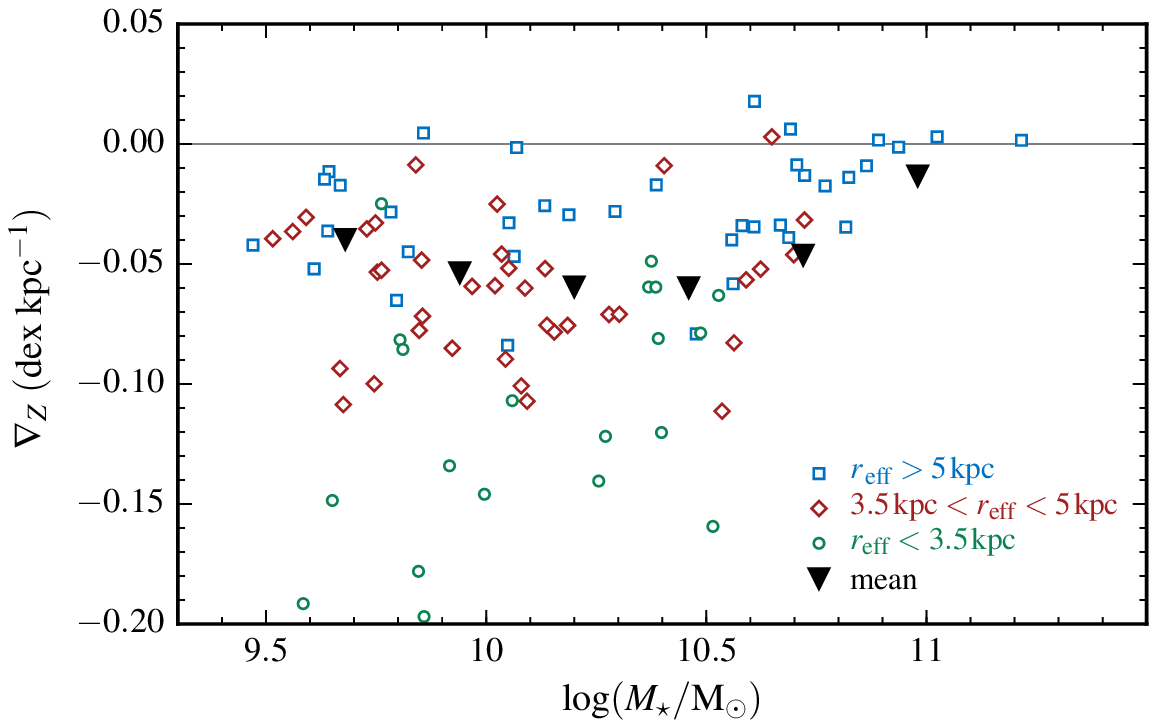}
\centering
\caption[]{Metallicity gradients as a function of stellar mass for all analysed redshifts. The colours denote different sizes of the discs $ r_{\rm eff} > 5$ kpc (blue stars),  $ 3.5 {\rm kpc } <r_{\rm eff} < 5$ kpc (red stars) and $ r_{\rm eff} < 3.5$ kpc  (black stars). 
Mean gradients for the stacked distribution are shown (black triangles).}\label{fig:metzreff}
\end{figure}

\subsection{Metallicity profiles of the stellar populations}
We estimate the radial abundance profiles of the SPs in the selected simulated discs for the four mentioned redshifts.
We perform linear regression fits by applying a bootstrap technique to the metallicity profiles.
In Fig. ~\ref{fig:metzmasa}, we show the stellar metallicity gradients as a function of stellar mass for the four analysed redshifts (upper panels). In 
the lower panels of this figure, the renormalised systems are displayed. As the minimum number of stellar particles are kept fixed as a function of redshift, the number of analysed discs decreases with increasing redshift.
The shaded areas denote the standard deviations.

The calculations of the metallicity gradients are done in the  range $[0.5, 1.5]\,r_{\rm eff}$\footnote{The $r_{\rm eff}$ is the half-mass radius of the disc component (i.e. excluding the bulge).} and  $[0.5, 1.0]\,r_{\rm eff}$, which  is the same interval used by \citet{tissera2016b,tissera2016a} for the analysis of the gas and stellar metallicity profiles at $z \sim 0$.
The results obtained by calculating the metallicity gradients in these two radial intervals are  in agreement with each other as can be seen from Table ~\ref{tab1}. We also tested extended  calculations to $[0.5, 2.0]\,r_{\rm eff}$. However, the latter encompasses the external
regions of the discs, which  introduce extra noise  when   the discs have breaks or nearby  satellite galaxies, and so on (Varela et al. in preparation),  or have lower numbers of stellar particles. 

We can see that, at all redshifts, metallicity gradients become steeper with stellar mass, but only up to  log M~$\sim 10.3$. For higher masses, there is an inversion of this trend, and metallicity gradients become flatter. This leads to a characteristic U-shape for the relation of the stellar metallicity gradient with the stellar mass.
The same behaviour is found for the normalised gradients. The overall trend  is not the expected one for a linear extrapolation of the relation for the high-mass galaxies (or vice-versa, for the extrapolation for the low-mass end).

The observational trends determined for galaxies in the CALIFA survey by  \citet{sanchezb2014}  are more comparable to estimations within  $[0.5, 1.5]\,r_{\rm eff}$. To illustrate this, in Fig. ~\ref{fig:metzmasa} we include  the observations by \citet{sanchezb2014}  and \citet{gonzalezdelgado2015} at $z\sim 0$. As reported by these papers,  there is trend for a change in slopes
so that  lower-stellar-mass galaxies have flatter metallicity gradients than expected from a single linear relation.
We note that a full agreement between the observed and the simulated normalised gradients is not expected since the scale-lengths and chemical abundances
are estimated differently. The important point to highlight is the inversion of the metallicity trend with stellar mass in the observed and simulated relations  at approximately the same stellar mass.

Previous works reported the existence of a correlation between gas-phase metallicity gradients and $r_{\rm eff}$  \citep[e.g.][]{prantzos2000,ho2015}, which could be understood
based on the hypothesis of angular momentum conservation of the gas as it flows in and settles into a disc structure.  
The correlation disappears when   the gas-phase metallicity gradients  are normalised by $r_{\rm eff}$ for both observed and simulated discs \citep[e.g.][]{sanchez2012,ho2015,tissera2016a}. 
Within the standard model for disc formation \citep{fall1980}, the sizes of the discs are the result of angular momentum conservation.
f We understand this process  
as the  net angular momentum conservation that can be the result of a balance between losses and gains along the evolutionary paths.
Considering the relation between the specific 
angular momentum and the stellar mass  reported for  observed and simulated disc galaxies \citep[e.g.][]{romanowsky2012, fall2013,pedrosa2014,genel2015, teklu2015}, the normalisation
by the  $r_{\rm eff}$ is expected to eliminate the mass-dependence, unless other processes act to strongly disturb the discs. Our results are in agreement with a scenario where  gaseous discs form, conserving the 
angular momentum, overall, and  the stellar populations form inside-out, enriching the ISM and producing  metallicity gradients.   These simulations take into account
the assembly in a hierarchical scenario and thus, perturbations to the discs due to inflows, interactions, or mergers are considered, as well as the action of galactic outflows driven by SN feedback.

  For the stellar metallicity gradients, \citet{tissera2016b}
reported a similar correlation at $z\sim 0$.  We extended these estimations to
up to $z \sim 1$, finding similar trends as shown in Fig. ~\ref{fig:metzradio}. The Pearson correlation factors for these
relations are 0.67, 0.61, 0.74, and 0.65 for $z \sim 0, 0.4, 1,$ and $ 1.3$, respectively.
For the normalised gradients, the Pearson factors are 0.51, 0.34, 0.63, and 0.44, respectively. Although the correlation signals
 are less significant for the normalised metallicity gradients,  they are high enough  to suggest the non-existence of  a common metallicity gradient for the 
SPs as the one found for the  gas-phase metallicity gradients in observations \citep{sanchez2014} and simulations \citep{tissera2016b}.

In order to gain  further insight into the distribution of metallicity gradients as a function of stellar mass, we look
for possible dependence on the global properties of the galaxies.  Interestingly,  the only clear trend is found 
with  $ r_{\rm eff}$.   In  Fig.~\ref{fig:metzreff}   we show all estimated metallicity gradients since $z \sim 1$ and the
median values of the stacked distribution. At a given stellar mass, disc galaxies
with flatter metallicity slopes  have
larger  $ r_{\rm eff}$. Disc galaxies with the more negative stellar metallicity gradients tend to have  the smaller  $ r_{\rm eff}$, on average.
As can be seen,   the inversion of the relation is recovered after the stacking of the four analysed redshifts. This suggests that the valleys of the 
distributions are located at approximately  the same stellar mass, regardless of analysed redshift. We might then expect  the physical process responsible 
for this behaviour to be independent of redshift. 

In previous works, it has been shown that  SN feedback is expected to
be more efficient at driving galactic winds in smaller galaxies. \citet{scan08} showed that the SN feedback model adopted in this work recovered such
behaviour, producing a more efficient regulation of the star formation activity and driving stronger mass-loaded galactic winds in galaxies with
shallower potential wells \citep[see also][]{derossi2010}.   As a consequence, disc components might be larger than expected since the gas can be heated up by SN feedback and blown out, acquiring angular momentum. Then it might cool down again contributing to build up the external parts of the discs where star formation can continue. This happens in some of the simulated discs within the lower
stellar mass range. It is for these galaxies that our simulations produce gas-phase disc components and stellar discs with larger variety in metallicity gradients. 
 Recent results on the stellar metallicity gradients from the MaNGA survey have been reported. \citet{goddard2017}  find most
stellar metallicity gradients to be negative in agreement with previous works  and to show  an anticorrelation with the stellar mass.  Using the MaNGA survey, \citet{zheng2017} reported a  correlation of the stellar metallicity gradients with stellar mass, albeit depending on galaxy morphology,  in agreement with results from the 
CALIFA survey  \citep{sanchezb2014, gonzalezdelgado2015}. Hence, observations do not  yet provide a clear description of how the metallicity of the
stellar populations  depends on  the stellar mass.

\subsection{The influence of radial migration}

\begin{figure}
\includegraphics[width=0.8\columnwidth]{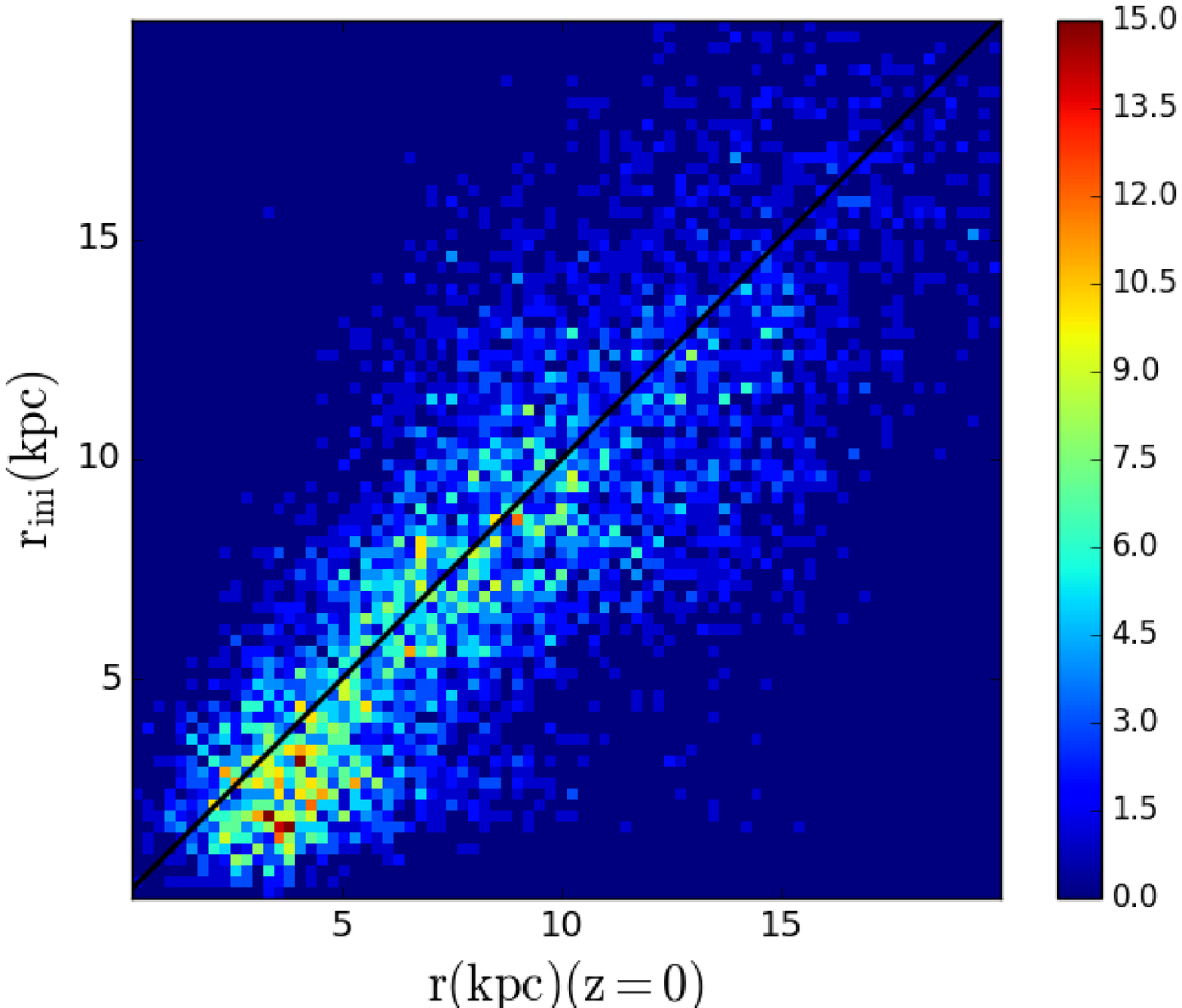}
\includegraphics[width=0.8\columnwidth]{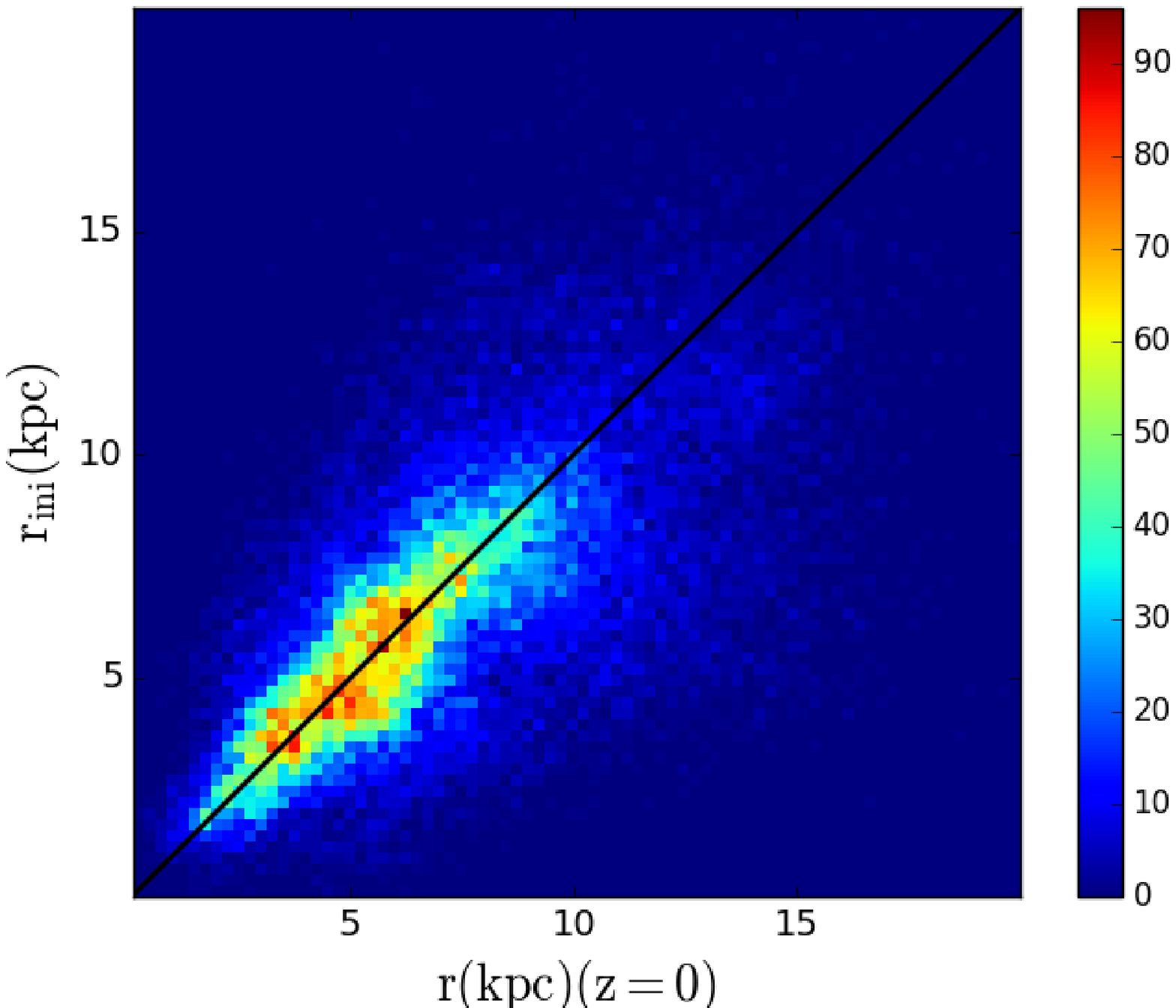}
\centering
\caption[]{ Initial radius (${\rm r_{in}}$) of stars (i.e. where it is first identified as a star) as a function of their radius (${\rm r_{(z=0)}}$) at $z\sim 0$ for a galaxy with strong (upper panel) and weak  (lower panel) signal of 
migration. Both galaxies have similar stellar metallicity gradients ($-0.02~{\rm dex kpc^{-1}}$ and  $-0.03 ~{\rm dex kpc^{-1}}$ , respectively) and similar stellar masses ($1.1\times 10^{10}{\rm M_\odot} $ and  $4.7\times 10^{10}{\rm M_\odot} $, respectively).}
\label{fig:migration}
\end{figure}

Migration is a process that could also contribute to flatten the stellar metallicity gradients in the disc components of galaxies \citep[e.g.][]{sellwood2002}. Our simulations do not allow us to study this effect in detail  as done by \citet{grand2016}, for example,   because of the limited number of available snapshots and the numerical resolution. Nevertheless, we made a rough estimation of the radial displacement of star particles. 
Following \citet{sanchezb2009}, in Fig.~\ref{fig:migration}, we show the relation between the initial radius and the radial position at $z\sim 0$ for two disc galaxies with similar stellar masses and metallicity gradients  (note that the initial radius corresponds to the first time the particle is identified as a star one). As can be seen there is variation in these correlations: one of them shows
 signs of  having experienced stronger migration in the outer parts (upper panel) while the next one shows  a larger displacement within the central regions (lower panel).  For these galaxies, 
$\sim 22$ per cent of the stars within 1.5 of the optical radius show  a displacement larger than $20$ percent of the initial radius. 
Overall, we measured up to $\sim 20-25$ per cent variation in the positions within the $1.5 r_{\rm eff}$ and the break radius (i.e. the radius where the stellar surface density showed a change of slope if a double power law is fitted). For this estimation, the analysis was done for the 16 most massive galaxies. This mass range goes down
to $\sim 10^{10.5}$.  In this mass interval, we did not find a clear trend with stellar mass, which could be associated to the inverse of the  the metallicity gradient relation (Fig.~\ref{fig:metzmasa}). Indeed, our galaxy sample does not cover the desired mass range with enough numerical resolution to evaluate this. 
Compared with the results of \citet{sanchezb2009} who found $\sim 40$ per cent displacement of stellar particles in the outskirts of a disc galaxy, our simulated galaxies show  weaker migration effects as mentioned above. 
We note that the SN feedback is different between the two models. Our numerical mass resolution  is   slightly better  than that used by  \citet{sanchezb2009} but \citet{tissera2016a} have adopted a larger gravitational softening,  which leads to a spatial resolution of $\sim 700$ pc. However, \citet{grand2016}, where a more detailed
analysis is performed, have higher resolution. Such a detailed analysis cannot be done with our simulations. 

As an alternative assessment  of the global impact of migration, we  resort to the archeological estimation of the
metallicity gradients using mono-age SPs identified at $z\sim 0$. The comparison of the  actual metallicity gradients at a given redshift
with the metallicity gradients of the SPs with similar age allows  a global assessment of the effects of migration in our simulations.
A larger diference between the evolution estimated by these two different methods would suggest an important effect of migration on our simulated metallicity
profiles.

\subsection{The evolution of the stellar metallicity gradients}

\begin{figure}
\includegraphics[width=0.9\columnwidth]{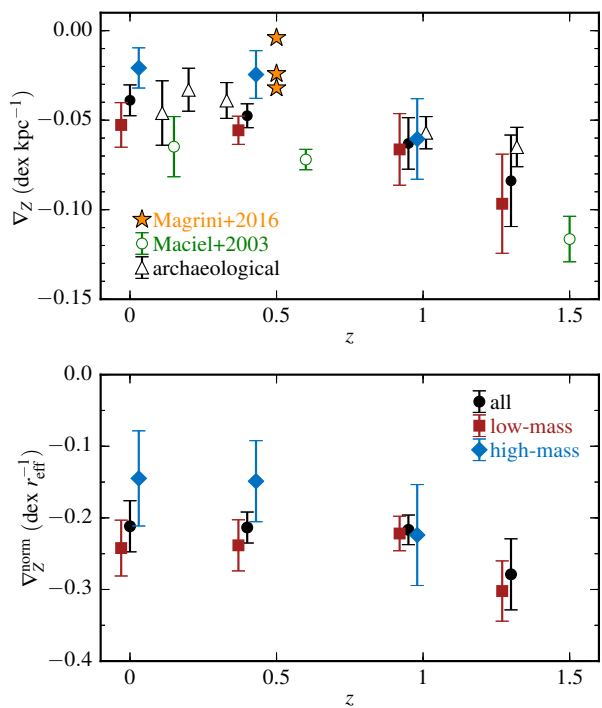}
\centering
\caption[]{Evolution of the stellar metallicity gradients. \textit{Upper panel:} Median slopes of the metallicity profiles for all simulated discs (black circles) and for those in the low-mass (red squares) and high-mass (blue diamonds) subsamples as a function of redshift. For comparison we include the observational estimations reported by \cite{maciel2003} for the Milky Way (green open circles) and by \citet{magrini2016} (orange stars) for nearby galaxies. Estimations of the median metallicity gradients (open triangles) performed using the ages and spatial distributions of stars at $z\sim 0$ are included (see text for details). \textit{Lower panel:} Similar estimations for the normalised metallicity gradients. In both panels, the error bars are estimated using a bootstrap technique. For the sake of clarity, the simulation points are slightly offset horizontally in both panels.}
\label{fig:metz}
\end{figure}

In Fig.~\ref{fig:metz} (upper panel) we show the median metallicity gradients as a function of redshift, for the whole galaxy sample and for the high- and low-stellar-mass subsamples.
We note that the median metallicity gradients of high-stellar-mass discs are systematically  flatter than those of systems in the low-stellar-mass subsamples for $z < 0.5$.
This trend can be seen from Table\ref{tab1}, where we summarise the metallicity gradients for their SPs as a function of stellar mass, for all analysed redshifts. 
However~, as our sample lacks massive galaxies at high redshift, it is difficult to distinguish between this effect and the actual change  in the metallicity gradients with stellar mass as a function of redshift. 
Hence, hereafter, we focus on the global evolution of the metallicity gradients. Globally,  we detect a mild evolution of the median metallicity gradients of the stellar discs in the analysed redshift interval as can be seen from Fig.~\ref{fig:metz} (upper panel) and Table 1. The mean change in slope is $-0.02 \pm 0.01$  dex~kpc$^{-1}$ from $z\sim 0 $ to $z\sim 1$, which is also consistent with no evolution. We note that for these estimations we are using the stellar  and metallicity distributions  of the disc galaxies at each analysed redshift.

A different approach is to follow the archeological method of defining  mono-age  SPs at $z=0$ and estimating  the metallicity gradients for each of them \citep[e.g.][]{maciel2003,magrini2016}. From those
metallicity gradients, it is possible to infer the level of evolution,  assuming stars do not strongly modify their spatial locations after their birth. For the MW, \citet{maciel2003} estimated a weak evolution  consistent with the metallicity gradients to be more
negative in the past. Recent estimations by \citet{magrini2016} for  galaxies in the Local Group suggest  no evolution since $z\sim 0.5$. These authors could not determine 
if the lack of evolution was  produced by a very effective SN feedback or by radial migration.

We apply the same method by defining mono-age SP adopting age intervals of $2$ Gyr.  In  Fig.~\ref{fig:metz} (upper panel),  we display
these median gradients (open triangles), which  are found to be in good agreement with those estimated using the actual distribution of the SPs in the discs at a given redshift (squares). The evolution derived from this method is $-0.03 \pm 0.02$ dex~kpc$^{-1}$ over the whole redshift range analysed.
For $z<0.5$, this  analysis predicts no evolution of the stellar metallicity gradient, on average.
 
The archeological estimations  encompass the effects of processes that can affect the distribution and dynamics of baryons such as galaxy interactions, strong bars, or migration.
 The fact that, on average, the evolution signals obtained from the stellar metallicity gradients of the stellar  populations at a given redshift 
and from those derived by archeological means are in agreement supports a mild impact of these effects on the evolution of stellar metallicity gradients.
 As  shown in this figure, our results are consistent with similar {results reported} for the MW stellar disc by \citet[][open circles]{maciel2003} and those  given by \citet{magrini2016} as shown above. We  note that these are global trends and as can be seen from Table~\ref{tab1} and  Fig.~\ref{fig:metz},   the dispersion is high enough to provide room
for a diversity due to the different history of assembly of the galaxies in a hierarchical clustering scenario.

%------------------------------------------------------------
%------------------------------------------------------------

%--------------------------------------------------------------------

\begin{table*}
\caption{ Median stellar metallicity gradients in ${\rm dex~kpc^{-1}}$ and ${\rm dex~{\it r}_{eff}^{-1}}$ for all, low and high stellar-mass galaxies in the
$z = [0, 1.3]$ interval. Estimations in both radial intervals are shown for completitude: $[0.5, 1.5]\,r_{\rm eff}$ and $[0.5, 1.0]\,r_{\rm eff}$. Bootstrap errors and standard deviations (parentheses) are included. Two sets of results are given, where the gradients were measured in different radial ranges.}
\label{tab1}
\centering
\begin{tabular}{c c c c c c}
\hline\hline
     &              &  \multicolumn{2}{c}{$[0.5\,r_{\rm eff}:1.0\,r_{\rm eff}]$}& \multicolumn{2}{c}{$[0.5\,r_{\rm eff}:1.5\,r_{\rm eff}]$} \\ 
%\cmidrule(r{5pt}){3-4} \cmidrule(l{5pt}){5-6}
$z$  & $M_{\star}$  & $\nabla_{Z}$           & $\nabla_{Z}$                     & $\nabla_{Z}$           & $\nabla_{Z}$                     \\
     & $M_{\odot}$  & $({\rm dex~kpc^{-1}})$ & $({\rm dex~{\it r}_{eff}^{-1}})$ & $({\rm dex~kpc^{-1}})$ & $({\rm dex~{\it r}_{eff}^{-1}})$ \\
\hline

\multirow{3}{*}{$ z = 0.0 $} & $ < 10^{10.5} $ & $-0.04 \pm 0.01~(0.04)$ & $-0.21 \pm 0.07~(0.18)$ &  $-0.05 \pm 0.01~(0.03)$ & $-0.24 \pm 0.04~(0.13)$  \\
                             & $ > 10^{10.5} $ & $-0.03 \pm 0.02~(0.05)$ & $-0.26 \pm 0.10~(0.27)$ &  $-0.02 \pm 0.01~(0.03)$ & $-0.15 \pm 0.07~(0.16)$  \\
                             & all masses      & $-0.04 \pm 0.01~(0.04)$ & $-0.23 \pm 0.07~(0.21)$ &  $-0.04 \pm 0.01~(0.04)$ & $-0.21 \pm 0.04~(0.15)$  \\
\\
\multirow{3}{*}{$ z = 0.4 $} & $ < 10^{10.5} $ & $-0.06 \pm 0.01~(0.03)$ & $-0.28 \pm 0.04~(0.14)$ &  $-0.06 \pm 0.01~(0.02)$ & $-0.24 \pm 0.04~(0.10)$  \\
                             & $ > 10^{10.5} $ & $-0.03 \pm 0.01~(0.03)$ & $-0.19 \pm 0.08~(0.17)$ &  $-0.02 \pm 0.01~(0.02)$ & $-0.15 \pm 0.06~(0.12)$  \\
                             & all masses      & $-0.05 \pm 0.01~(0.03)$ & $-0.25 \pm 0.04~(0.16)$ &  $-0.05 \pm 0.01~(0.03)$ & $-0.21 \pm 0.02~(0.12)$  \\
\\
\multirow{3}{*}{$ z = 0.95 $} & $ < 10^{10.5}$ & $-0.07 \pm 0.03~(0.12)$ & $-0.23 \pm 0.09~(0.31)$ &  $-0.07 \pm 0.02~(0.07)$ & $-0.22 \pm 0.02~(0.14)$  \\
                              & $ > 10^{10.5}$ & $-0.05 \pm 0.04~(0.11)$ & $-0.15 \pm 0.10~(0.25)$ &  $-0.06 \pm 0.02~(0.05)$ & $-0.22 \pm 0.07~(0.14)$  \\
                              & all masses     & $-0.05 \pm 0.02~(0.11)$ & $-0.19 \pm 0.06~(0.29)$ &  $-0.06 \pm 0.01~(0.06)$ & $-0.22 \pm 0.02~(0.14)$  \\
\\
\multirow{3}{*}{$ z = 1.3 $} & $ < 10^{10.5}$  & $-0.08 \pm 0.02~(0.13)$ & $-0.25 \pm 0.08~(0.30)$ &  $-0.10 \pm 0.03~(0.08)$ & $-0.30 \pm 0.04~(0.14)$  \\
                             & $ > 10^{10.5}$  &     --                  &     --                  &      --                  &     --                   \\
                             & all masses      & $-0.07 \pm 0.03~(0.13)$ & $-0.21 \pm 0.08~(0.31)$ &  $-0.08 \pm 0.03~(0.08)$ & $-0.28 \pm 0.05~(0.17)$  \\

\hline
\end{tabular}
\end{table*}

 Our findings support  the inside-out formation scenario (driven by angular momentum conservation) as  the main process at work for spiral galaxies.  However, this would  not prevent other mechanisms from taking place and mixing up the SPs such as bars or migration. According to our results, they might be secondary effects which contribute to flattening the metallicity slopes  and to increasing the dispersion depending on the history of formation
of each galaxy (i.e. the formation of bars, subsequent gas accretion, or galaxy encounters). 

%%%%%%%%%%%%%%%%%%%%%%%%%%%%%%%%%%%%%%%%%%%%%%%%%%%%%%%%%%%
\section{Conclusions}
\label{sec:conclu}

We study  the evolution of the metallicity gradients of the stellar discs in a hierarchical clustering scenario with the aim of analysing their evolution with time. The metallicity gradients reflect the enrichment of the ISM at the time of star formation and hence, can store information relevant to understanding the star formation process, the enrichment cycle, and the effects of galaxy assembly. Later, stars can be redistributed in the discs via dynamical processes such as  galaxy-galaxy  interactions or migration, altering the chemical patterns. Hence, the evolution of the stellar metallicity gradients holds important information for understanding galaxy formation.

We find a trend for our stellar discs to have  metallicity gradients consistent with  an inversion of the relation as a function of stellar mass. Our results show  that some  galaxies with stellar masses smaller than  $\sim 10^{10.3}M_\odot $  have  flatter stellar metallicity
gradients  than expected from a linear correlation.  These galaxies are systematically found to have larger sizes than those with steeper negative metallicity slopes at a given stellar mass.  As a consequence, a common metallicity gradient for the  SPs as a function of stellar mass cannot be determined.   
SN feedback might be a major process determining the shape of the
stellar metallicity versus stellar mass distributions, as it is known to be stronger in galaxies with shallower potential wells. The impact of the SN feedback on the metallicity distribution also depends on the history of assembly galaxies. 

The metallicity gradients (dex ${\rm kpc^{-1}}$)  correlate statistically better with  $r_{\rm eff}$. This relation  also shows a clear increase of the dispersion for lower masses.
The normalisation of the stellar metallicity gradient-size relation does not  erase the correlation as  is case for the gas-phase metallicity gradients \citep{tissera2016a}.
The action of dynamical effects could modify  the metallicity gradients after the gas is transformed into stars. In the case of  
the gaseous discs, continuous gas infall can contribute to keeping such a relation in place, if it occurs with angular momentum conservation.
Some of the simulated discs show evidence for migration; the estimated fractions are in the range $[10,25]$ per cent. 
 These fractions have been robustly estimated for galaxies with stellar masses larger than  $\sim 10^{10.3} {\rm M_{\odot}}$. However, a similar analysis for smaller galaxies
could not be done due to limited numerical resolution.

The stellar metallicity gradients are found to mildly evolve by $-0.02 \pm 0.01~{\rm dex~kpc^{-1}} $ from $z\sim 0 $ to $z \sim 1$, albeit with a larger dispersion. This evolution has been obtained by using the stellar and metallicity distributions in each analysed disc at a given redshift.
The normalisation of metallicity gradients by the corresponding $r_{\rm eff}$ of each stellar disc component provides a median slope, which does not vary statistically up to $z \sim 1$. We note that there is a large dispersion at each analysed redshift, which can be ascribed to the 
variety of evolutionary paths. For  $z > 1$ we find indications of stronger evolution that should be constrained with a larger galaxy sample. 

We also perform archaeological estimations of the metallicity gradients defining  mono-age SPs. These yield an evolution of $-0.03 \pm 0.02 ~{\rm dex~kpc^{-1}}$ , which is in agreement with that estimated from the actual distribution of stars. This finding supports a weak effect of stellar migration in our simulations, on average.
We also point
out that the level of evolution of the stellar metallicity gradients predicted by our simulations is consistent with those reported by recent observations  that used the archeological approach \citep{maciel2003,magrini2016}. 

Our results suggest that SN feedback  is one  mechanism responsible of
producing more extended stellar discs with shallower metallicity gradients than expected for the extrapolation of the  linear correlation with stellar mass. Migration could also explain the  flattening of the  metallicity
gradients if it acts preferentially in some galaxies; for example, if they have experienced strong bar formation \citep{grand2016}.  Furthermore, galaxy-galaxy interactions or small satellite accretions can also modify the metallicity profiles in the outer parts \citep[e.g.][]{quillen2009,perez2011}. Disentangling the effects of these processes  remains a challenge
in a cosmological context.

%%%%%%%%%%%%%%%%%%%%%%%%%%%%%%%%%%%%%%%%%%%%%%%%%%%%%%%%%%%
\begin{acknowledgements}

{We thank the anonymous referee for useful comments and suggestions.} This work was partially supported by PICT 2011-0959 and PIP 2012-0396 (Mincyt, Argentina). PBT acknowledges partial support from the Nucleo UNAB 2015 DI-677 of Universidad Andres Bello and Fondecyt 1150334 and the Southern Astrophysics Network (SAN) collaboration funded by Conicyt. REGM acknowledges support from \textit{Ci\^encia sem Fronteiras} (CNPq, Brazil).
PSB acknowledges financial support from the: CONICYT-Chile Basal-CATA PFB-06/2007 and the AYA2013-48226-C3-1-P by the Ministerio de Ciencia e Innovacion.
JMV acknowledges financial support from
AYA2013-47742-C4-1-P by the Spanish MINECO.

\end{acknowledgements}

%%%%%%%%%%%%%%%%%%%%%%%%%%%%%%%%%%%%%%%%%%%%%%%%%%%%%%%%%%%
\bibliographystyle{aa}
\bibliography{28915JN}

\end{document}